\documentclass{aa}

\usepackage{ulem}

\usepackage{lscape}
\usepackage{natbib}
\usepackage{hyperref}
\usepackage{txfonts}
\usepackage{graphicx}

\def\XMM{{\it XMM-Newton}}
\def\ROSAT{{\it ROSAT}}
\def\SourceHP{[HP99]~456}
\def\SourceXMM{XMMU\,J0541.8-6659}
\newcommand{\SII}{[S\,{\sc ii}]}
\newcommand{\OIII}{[O\,{\sc iii}]}
\newcommand{\Halpha}{H${\alpha}$}

\begin{document}
\normalem

\title{XMMU\,J0541.8-6659, a new supernova remnant\\ in the Large Magellanic Cloud\thanks{Based on observations with \XMM, an ESA Science Mission with instruments and contributions directly funded by ESA Member states and the USA (NASA).}}

\author{
M.-H.~Grondin\inst{1}, 
M.~Sasaki\inst{1},
F.~Haberl\inst{2},
W.~Pietsch\inst{2},
E.~J.~Crawford\inst{3}, 
M.~D.~Filipovi\'c\inst{3},\\
L.~M.~Bozzetto\inst{3},
S.~Points\inst{4},
R.~C. Smith\inst{4}
}
\institute{Institut f$\rm\ddot{u}$r Astronomie und Astrophysik T$\rm\ddot{u}$bingen, Universit$\rm\ddot{a}$t T$\rm\ddot{u}$bingen, Sand 1, 72076 T$\rm\ddot{u}$bingen, Germany 
\and Max-Planck-Institut f$\rm\ddot{u}$r extraterrestrische Physik, Giessenbachstrasse, 85748 Garching, Germany
\and University of Western Sydney, Locked Bag 1797, Penrith South DC, NSW 1797, Australia
\and Cerro Tololo Inter-American Observatory, Casilla 603, La Serena, Chile}
\titlerunning{XMMU\,J0541.8-6659, a new SNR in the LMC}
\authorrunning{Grondin et al., 2011}

  \abstract
{The high sensitivity of the \XMM{} instrumentation offers the opportunity to study faint and extended sources in the Milky Way and nearby galaxies such as the Large Magellanic Cloud (LMC) in detail. The \ROSAT\ PSPC survey of the LMC has revealed more than 700 X-ray sources, among which there are 46 supernova remnants (SNRs) and candidates.}
{We have observed the field around one of the most promising SNR candidates in the \ROSAT\ PSPC catalogue, labelled \SourceHP{} with \XMM{}, to determine its nature.}
{We investigated the \XMM{} data along with new radio-continuum, near infrared and optical data. In particular, spectral and morphological studies of the X-ray and radio data were performed.}
{The X-ray images obtained in different energy bands reveal two different structures. Below 1.0 keV the X-ray emission shows the shell-like morphology of an SNR with a diameter of $\sim$~73~pc, one of the largest known in the LMC. For its thermal spectrum we estimate an electron temperature of (0.49~$\pm$~0.12)~keV assuming non-equilibrium ionisation. The X-ray images above 1.0 keV reveal a less extended source within the SNR emission, located ~1\arcmin\ west of the centre of the SNR and coincident with bright point sources detected in radio-continuum. This hard component has an extent of 0.9\arcmin\ (i.e. $\sim$~13 pc at a distance of $\sim$~50~kpc) and a non-thermal spectrum. The hard source coincides in position with the \ROSAT\ source \SourceHP{} and shows an indication for substructure.}
{We firmly identify a new SNR in the LMC with a shell-like morphology and a thermal spectrum. Assuming the SNR to be in the Sedov phase yields an age of $\sim$~23 kyr. We explore possible associations of the hard non-thermal emitting component with a pulsar wind nebula (PWN) or background active galactic nuclei (AGN). }
\keywords{galaxies: Magellanic Clouds, ISM: supernova remnants}

\maketitle

\section{Introduction}

The study of supernova remnants (SNRs) is crucial for a complete understanding of the chemical composition and evolution of the ISM in a galaxy because of their energy and matter inputs into the interstellar medium (ISM). The sample of SNRs studied in our Galaxy is biased because of the high absorption in the Galactic disk. Therefore, observations of nearby galaxies allow us to perform unbiased population studies and constrain the physical properties of the sources in detail. It is thus also possible to better understand the evolution and structure of the ISM in the Milky Way.

Located at a distance of $\sim50$ kpc to the Earth \citep{Freedman2001, Macri2006}, the Large Magellanic Cloud (LMC) offers the ideal laboratory for studying a large sample of different types of objects (such as SNRs) in greater detail than in any other galaxy. Since its first detection in X-rays \citep{Mark1969}, the LMC has been extensively observed, but the major step forward came from more than 200 observations in a $10^{\circ} \times 10^{\circ}$ field centred on the LMC, which have been performed with the \ROSAT\ Position Sensitive Proportional Counter (PSPC) from 1990 to 1994. For a description of the \ROSAT\ mission and PSPC detectors, see \cite{Trumper1982}, \cite{Briel1986} and \cite{Pfeffermann1987}. This survey revealed 758 sources \cite[hereafter labelled HP99]{HP99}, among which 46 sources were classified as firmly identified SNRs or candidates. 

Several SNRs in the LMC have been investigated using observations with \ROSAT\ \citep{1998A&AS..127..119F, Williams1999}, {\it Chandra} X-ray Observatory \citep{Hughes2006, Seward2010} and \XMM{} \citep{Williams2004, Klimek2010, Crawford2010} satellites, allowing a more detailed view of their morphologies and spectra.  \citet{Badenes2010} have studied the size distribution of the SNRs in the Magellanic Clouds (MCs), which has a maximum at $\sim$40~pc and may extend up to sizes of $\sim$100~pc. With an extent of over 100~pc, SNR~0450-70.9 and SNR~0506-6542 (DEM~L\,72) are among the largest SNRs detected in the LMC \citep{Williams2004, Cajko2009, Desai2010, Klimek2010}, which may be highly evolved (age up to 100 ~kyr). The size distribution of the MC SNRs as well as those in our Galaxy or the neaby spiral galaxy M\,33 cannot be explained only by the Sedov expansion model for SNRs, but seems to be largely affected by the ambient ISM densities \citep[][and references therein]{Bandiera2010, Badenes2010}.

Furthermore, multi-frequency observations of several SNR candidates located in the LMC have enabled their firm identification based on morphological and spectral criteria \citep{Bojicic2007, Crawford2010} and have revealed a strong correlation between the X-ray sources and the emission observed by the Magellanic Cloud Emission Line Survey \cite[MCELS;][]{Smith2000}. Indeed, an enhancement of the \SII\ and \Halpha{} coincident with the X-ray emission can be observed in most cases. In particular, the ratio \SII/\Halpha{} is often higher than 0.4 \citep{Levenson1995, Williams2004}. However, several SNRs in the Magellanic Clouds such as SNR~LMC~J0528-6714 or/and SMC~SNR~J010505-722319 do not have any optical emission. 

While the emission of the gas shocked by the shock waves of SNRs is mainly of thermal nature, there can also be a pulsar or a pulsar wind nebula (PWN) in a SNR, which produce non-thermal emission. Pulsars are rapidly rotating neutron stars characterised by short periods (up to a few 10~s) and high surface magnetic fields \citep{Manchester2005}. The dissipation of the rotational energy of pulsars via magnetised particle winds can be at the origin of PWNe \citep{Gaensler2006}. A high percentage of PWNe known in our Galaxy have been detected in X-rays, and present a power-law spectra with a mean spectral index of $\Gamma$~=~--1.8~$\pm$~0.6 \citep{Kargaltsev2008, Kargaltsev2010}. Sensitive X-ray observations have enabled the detection of several PWNe and candidates within the Magellanic Clouds, with similar properties as PWNe in the Milky Way \citep{Gaensler2003, Williams2005, Owen2011}.

There are now over 50 well-established SNRs in the LMC \citep[][and references therein]{Badenes2010, Klimek2010} and some additional $\sim$~20 SNR candidates \citep{Bozzetto2011}. 
This would comprise one of the most complete samples of SNRs in external galaxies. Therefore, it is of prime interest to study LMC SNRs and compare them with SNRs in other galaxies such as M\,33 \citep{Long2010}, M\,83 \citep{Dopita2010}, the Small Magellanic Cloud \citep[SMC; ][]{Filipovic2005, Payne2007, Filipovic2008} and our Galaxy \citep{Stupar2008, Green2009}.

The SNR candidates in the \ROSAT\ PSPC catalogue have been classified based on the X-ray spectrum and spatial extent. Additional comparison to radio data taken with the Molonglo Observatory Synthesis Telescope \cite[MOST; $\nu$=843~MHz; ][]{Turtle1984} and with optical data of MCELS has shown that there are ROSAT PSPC sources with radio or optical counterparts indicative of an SNR but with a hard X-ray spectrum and therefore no typical SNR characteristics in X-rays. The source \SourceHP{} is one of the most promising new SNR candidates of this kind, with hard emission detected by \ROSAT\ PSPC and a possible radio or optical counterpart. It has been recently re-investigated through new observations with \XMM{} and the Australia Telescope Compact Array \cite[ATCA; ][]{Hughes2007}. 
 
In this paper, we report on the results of the analysis of new \XMM{} and ATCA follow-up observations of the source \SourceHP{}. Section~\ref{section:observations} presents the observations and analysis techniques. Results of the multi-frequency analyses are presented in Section~\ref{section:analysis}. Section~\ref{section:discussion} presents the discussion on the SNR properties and on the different scenarios to explain the non-thermal emission. In particular, we explore a possible association of this second component with a PWN or an AGN.  Conclusions are presented in Section~\ref{section:conclusions}.

\section{Observations and data reduction}
 \label{section:observations}

\subsection{X-rays}
 \label{datareduction_Xray}

The \XMM{} satellite is an X-ray observatory operated by the European Space Agency (ESA). The source \SourceHP{} has been proposed for observations with \XMM{} (Obs. Id. 0651880101, P.I. : M. Sasaki). This paper presents results of the X-ray analysis of a 20~ks observation obtained on this source. 

The source \SourceHP{} has been observed on 2010 June 06 (from 04:18:56 to 09:52:50 UT) with the European Photon Imaging Camera (EPIC) in full-frame mode and thin filters. Using the EPIC MOS1, MOS2 and pn CCDs, it offers the opportunity to perform sensitive X-ray observations of a field of the sky of diameter of 30\arcmin. More detailed technical descriptions of the EPIC cameras are presented by \cite{Turner2001} and \cite{Struder2001}.

The EPIC data were analysed with SAS v10.0.0\footnote{Science Analysis Software: http://xmm.esac.esa.int/sas/.}. The exposure time, after removing periods of high background, is $\sim$~11~ks. Pixels flagged as bad were not taken into account and screening on the patterns (from 0 to 12 for MOS; from 0 to 4 for pn), corresponding to the canonical set of valid X-ray events (calibrated on the ground) was applied for the image and spectral analyses. We performed source detection by using the SAS tasks {\it eboxdetect} and {\it emldetect}. 
\begin{figure}[t!!]
\begin{center}
\includegraphics[width=0.48\textwidth]{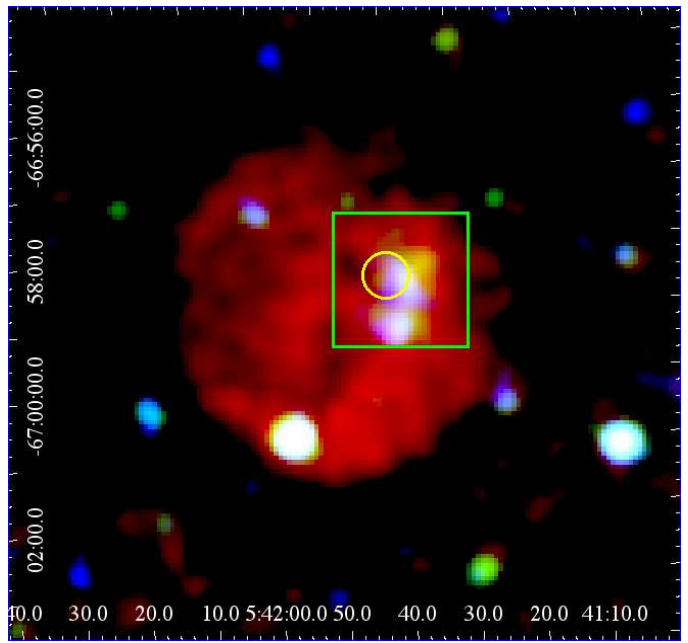}
\caption{Smoothed RGB three-colour image of combined exposure-corrected \XMM{} EPIC pn, MOS1 and MOS2 data (red: 0.2 - 1.0 keV; green : 1.0 - 2.0 keV; blue : 2.0 - 4.5 keV, square root scale). The instrumental background was estimated using the method described in Section \ref{subsection:SpectrumBackground} and subtracted from the images. The renormalisation factor was derived from the shaded detector corners. The field of the radio image presented in Figure~\ref{fig:figure.radio2} is overlaid for comparison (green square). The position of the \ROSAT\ source [HP99] 456 is represented by a yellow circle.}
\label{fig:figure1}\label{fig:figure1bis}
\end{center}
\end{figure}

\subsection{Radio-continuum}
 \label{datareduction_radio}

The field of \SourceHP{} was observed with the Australia Telescope Compact Array (ATCA) on 2010 November 29, with an array configuration 6C, at wavelengths of 3 and 6~cm (9000 and 5500~MHz), and a bandwidth of 2~GHz (ATCA project C2367). The observations were carried out in snap-shot mode, totalling about 1 hr of integration over a 12 hr period. PKS B1934-638 was used for flux and bandpass calibration and PKS 0530-727 was used for phase calibration. Standard calibration, editing and imaging techniques \citep{Sault2011} were used. Large bandwidth multifrequency clean \citep{Sault1994} was used to deconvolve the image. We point out that interferometers such as the ATCA suffer from missing flux owing to the lack of short spacings, which significantly affects the overall detection of extended emission like that from SNRs.

We also used various other radio observations (See Table~\ref{tab-flux}) including 843~MHz by \citet{Mills1984}, 1377~MHz from \citet{Hughes2007} and 4800~MHz \& 8640~MHz taken from the mosaic project of \citet{Dickel}.

\subsection{Optical}
 \label{datareduction_optical}

The Magellanic Cloud Emission Line Survey (MCELS)\footnote{MCELS : http://www.ctio.noao.edu/mcels/} was carried out from the 0.6~m University of Michigan/CTIO Curtis Schmidt telescope, equipped with a SITE $2048 \times 2048$\ CCD, which gave a field of 1.35\degr\ at a scale of 2.4\arcsec\,pixel$^{-1}$ \citep{Smith2006, Winkler2011}. Both the LMC and SMC were mapped in narrow bands corresponding to \Halpha, \OIII\ ($\lambda$=5007\,\AA), and \SII\ ($\lambda$=6716,\,6731\,\AA), plus matched red and green continuum bands. All data were continuum subtracted, flux-calibrated and assembled into mosaic images (a small section of which is shown in Figure~\ref{fig:figureHa}).


\section{Data analysis}
 \label{section:analysis}

\subsection{X-ray morphology}
 \label{subsection:Morpho}

The morphological analysis was performed in three energy ranges : 0.2~--~1.0 keV (soft band), 1.0~--~2.0 keV (medium band), 2.0~--~4.5 keV (hard band). For each energy interval, exposure-corrected images were obtained using SAS tools using the following method. 

First, X-ray data were cleaned using the selection procedure described in the previous section. We computed event images and exposure maps using the filtered dataset. Then we applied a mask to remove bad pixels from the three instruments. We divided each resulting event image by the corresponding exposure map and smoothed the resulting images with a Gaussian filter using {\it asmooth}. Finally, we added the event images from the different instruments for each energy band.

Figure~\ref{fig:figure1} shows the RGB exposure-corrected image around the unidentified source \SourceHP{} obtained with the EPIC data in the three bands defined above (red: 0.2~--~1.0~keV; green : 1.0~--~2.0~keV; blue : 2.0~--~4.5~keV) after subtraction of the instrumental background (see caption). The emission seen in the hardest band is similar to the medium band. This morphological analysis reveals the existence of two distinct emitting regions:
 
\begin{enumerate}
\item A soft emitting region: below 1.0~keV, the X-ray emission is dominated by a bright thermal (for more details, see Section~\ref{subsection:Spectrum}) component. It presents a shell-like morphology with an extent of $\sim$~5.0\arcmin $\times$ 4.6\arcmin.

\item A hard emitting region: above 1.0 keV, the X-ray analysis reveals a less extended harder component. Its centre is located $\sim$~1\arcmin\ west from the centre of the soft shell-like emission and presents an elongated morphology of length $\sim$~0.9\arcmin. The position is consistent with the \ROSAT\ position of \SourceHP{}.
\end{enumerate}

The presence of the hard emitting region coincident with the \ROSAT\ source suggests that \SourceHP{} is unrelated to the shell-like emission region. Therefore, we designated the soft, shell-like source \SourceXMM{} according to its approximate central coordinates. Additional details on the positions and extents of both emitting regions are summarised in Table~\ref{tab:Morpho}. The count rates of the EPIC-pn observations are listed in the last column. 

\begin{figure}[ht!!]
\begin{center}
\includegraphics[width=0.48\textwidth]{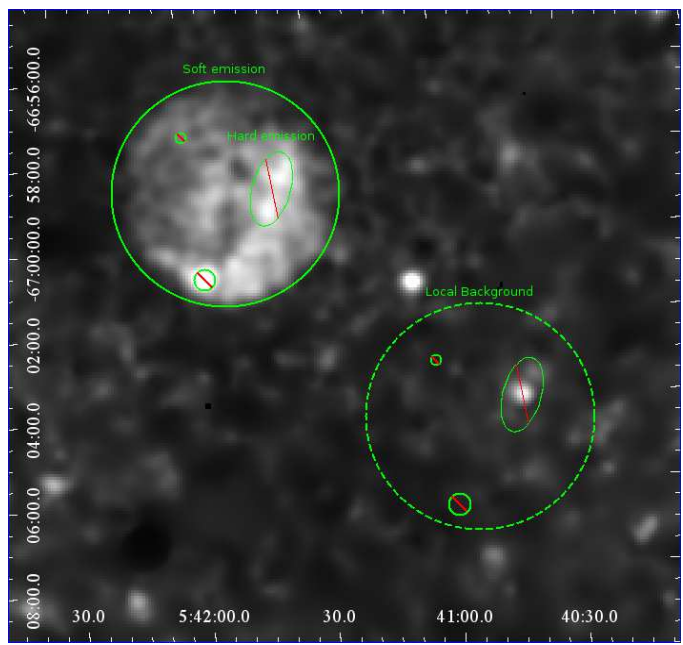}
\caption{Smoothed image of exposure-corrected XMM-Newton EPIC data in the 0.2~--~1.0 keV energy range. The dashed circle shows the region considered for the estimation of the local background. The thick circle represents the region used to derive the spectrum of the soft emission from the shell-like structure. Cancelled circles show the position of the bright sources excluded from the spectral analysis.}
\label{fig:backgroundregion}
\end{center}
\end{figure} 

\begin{table*}
\begin{center}
\caption{Integrated radio flux densities of \SourceHP{}.}
 \label{tab-flux}
\begin{tabular}{cccccl}
\hline\noalign{\smallskip}
\hline\noalign{\smallskip}
$\nu$ & $\lambda$ & R.M.S & Beam Size & S$_\mathrm{Total}$ & Reference \\
(MHz) & (cm)      & (mJy) & (\arcsec) & (mJy)              & \\
 \noalign{\smallskip}\hline\noalign{\smallskip}
 \hline\noalign{\smallskip}
843  & 36 & 1.5 & 43.0$\times$43.0 & 0.045    &   \citet{Mills1984} \\
1377 & 20 & 1.5 & 45.0$\times$45.0 & 0.029    &   \citet{Hughes2007} \\
4800 & 6  & 0.6 & 30.0$\times$30.0 & 0.008    &   \citet{Dickel}\\
8640 & 3  & 0.6 & 12.0$\times$12.0 & 0.004    &   \citet{Dickel} \\
5500 & 6  & 0.1 & 2.8$\times$2.2 & ---    &   This Work \\
9000 & 3  & 0.1 & 2.8$\times$2.2 & ---    &   This Work \\
 \hline\noalign{\smallskip}
\end{tabular}
\end{center}
\end{table*}

\begin{table*}
\centering
\caption{\label{tab:Morpho} Morphological details on the two emitting regions. The last column lists the count rates estimated from the analysis of EPIC-pn data between 0.3 and 5.0 keV.
} 
\begin{tabular}{lccccc}
\hline\hline 
Emitting & Ra & Dec & Size & Position angle & pn count rate\\
 region & (hh:mm:ss.d) & ($^{\circ}$:\arcmin:\arcsec) & (\arcmin) & ($^{\circ}$) &(cts~s$^{-1}$) \\
\hline
Soft emission & 05:41:51.5 & -66:59:02.8 & 5.0 $\times$ 4.6 & --45 & (18 $\pm$ 2)~$\times$ 10$^{-2}$\\
Hard emission & 05:41:39.4 & -66:58:45.8 & 0.9 $\times$ 0.45 & --15 & (2.8 $\pm$ 0.3)~$\times$ 10$^{-2}$ \\
\hline
\end{tabular}
\end{table*}


\subsection{X-ray spectral analysis}
 \label{subsection:Spectrum}

The following sections present the results of the subsequent spectral analyses of the two spectral components mentioned above. We first describe the method used to estimate the contribution from the intrinsic detector and X-ray background for each emitting region. The spectral analysis was performed using the XSPEC v~12.6.0 package \citep{Arnaud1996, Dorman2001}.

Point sources that were detected using the method described in Section~\ref{datareduction_Xray} were excluded from the spectral analysis.We used data between 0.3 keV and 5.0 keV because no significant emission is detected at higher energies.

\subsubsection{Estimation of the instrumental background}
 \label{subsection:SpectrumBackground}

We estimated the contribution from the intrinsic instrumental background using the filter wheel closed (FWC) data \citep{Freyberg2001a, Freyberg2001b}. The intrinsic background is composed of internal electronic noise as well as continuous and fluorescent X-ray emission induced by high-energy particles. This background is measured by operating the EPIC instruments in the ``closed'' filter wheel position, where no photons from astrophysical sources can be observed. The FWC spectrum needs to be renormalised using the continuum at higher energies.

\subsubsection{Estimation of the X-ray background}
 \label{subsection:SpectrumXBackground}

The contribution of the X-ray local background emission was derived from the same observation as the source itself. We defined a region located close to our source (see Figure~\ref{fig:backgroundregion}), and extracted the spectrum from the observation and the FWC data. We estimated the renormalisation factor of the FWC spectrum with respect to the observed spectrum using the total count rates measured above 5~keV in the corners outside the field of view of the EPIC camera. The value of the renormalisation factor is 1.3.

The spectrum of the diffuse emission after subtracting the instrumental background can be modelled as a sum of the following three components:
\begin{itemize}
\item a soft component, modelling the emission from the local bubble with a collisional plasma, non-equilibrium model \citep[NEI; ][]{Borkowski1994, Borkowski2001, Hamilton1983, Liedahl1995}\footnote{For more details on XSPEC models, please see http://heasarc.nasa.gov/xanadu/xspec/manual/manual.html} assuming a low temperature (0.1~keV);
\item a hard component modelled with an absorbed NEI model with a temperature of 0.3~keV, which stands for the emission from the Galactic halo;
\item an absorbed power-law of spectral index $\Gamma=1.46$, to describe the extragalactic unresolved emission.
\end{itemize}
The parameters are fixed to values similar to those used by \cite{Kuntz2010}.

The two last components are convolved with a T$\rm\ddot{u}$bingen-Boulder ISM absorption model (TBABS). The Galactic foreground hydrogen column density of $0.6 \times 10^{21}$~cm$^{-2}$ was derived from \cite{Stark1992} and was used for the absorption in the Galaxy.

\begin{table}[b]
\centering
\caption{\label{tab:Spectrum} XMM-Newton spectral results of the soft thermal emitting region in the 0.3 - 5.0 keV energy range. Model is VNEI with abundances fixed to 0.5 of the solar abundances.} 
\begin{tabular}{lc}
\hline\hline 
Parameter	& Value\\
\hline
N$_H$ ($\times$10$^{21}$ cm$^{-2}$) in the Galaxy & 0.6 (fixed) \\
N$_H$ ($\times$10$^{21}$ cm$^{-2}$) in the LMC & 0 (fixed) \\
kT (keV) & 0.49 $\pm$ 0.12 \\
Ionisation timescale* $\tau$ (s~cm$^{-3}$) & (1.5 $\pm$ 0.7) $\times$ 10$^{10}$ \\
Absorbed flux (0.3 -- 5.0 keV, erg/cm$^2$/s) & $2.5 \times 10^{-13}$ \\
Absorbed luminosity (0.3 -- 5.0 keV, erg/s) & $5.1 \times 10^{34}$ \\
Unabsorbed luminosity (0.3 -- 5.0 keV, erg/s) & $8.9 \times 10^{34}$  \\
Reduced $\chi^2$ & 1.27 \\
Degrees of freedom & 86\\
\hline
\multicolumn{2}{l}{*: The ionisation timescale $\tau$ = n$_e$ t, where n$_e$ is the electron number }\\
\multicolumn{2}{l}{density and t is the age of the gas.}
\end{tabular}
\end{table}

\subsubsection{Soft emission region}
 \label{subsection:SpectrumShell}

The X-ray emission below 1.0 keV is dominated by a structure presenting a shell-like morphology with a maximal diameter of $\sim$~5.0\arcmin\  and centred at RA(J2000)~=~05$^h$41$^m$51.5$^s$, DEC(J2000)~=~--66\degr59\arcmin03.8\arcsec. Photons located within the region corresponding to the hard X-ray emitting region, as defined in Table~\ref{tab:Morpho}, were excluded from the source spectrum to avoid any contamination, as were detected point sources visible in Fig.~\ref{fig:figure1}. 

The spectrum of the shell-like structure was obtained after the subtraction of the instrumental background. The X-ray background was estimated using a local background spectrum extracted from the same data. The source spectrum and the local background were modelled simultaneously in XSPEC by using a background model consisting of the components as explained in the previous section. All parameters of the three background components were fixed except the normalisations. The normalisations of the background components and the parameters of an additional model component for the shell emission were free fit parameters.

The spectral analysis of the soft circular emitting region reveals a thermal spectrum. It can be modelled with a single-temperature non-equilibrium ionisation collisional plasma model  \citep[VNEI; ][]{Borkowski1994, Borkowski2001, Hamilton1983, Liedahl1995} with a temperature of (0.49~$\pm$~0.12)~keV with an absorption of N$_H$~=~0.6~$\times$~10$^{21}$~cm$^{-2}$ for the Galactic foreground (with solar abundances) and a value of  N$_H$ for the absorption in the LMC. The abundances were fixed to 0.5 of the solar system abundances for the emission and absorption taking place in the LMC. This value is the standard mean value for the LMC \citep{RussellDopita1992}. The value of N$_H$ for the LMC turned out to tend towards 0 with an 90~\% C.L. upper limit of 0.4~$\times$~10$^{21}$~cm$^{-2}$. This parameter was therefore fixed to 0. 

\begin{figure}[ht!!]
\begin{center}
\includegraphics[width=0.37\textwidth, angle=-90.0]{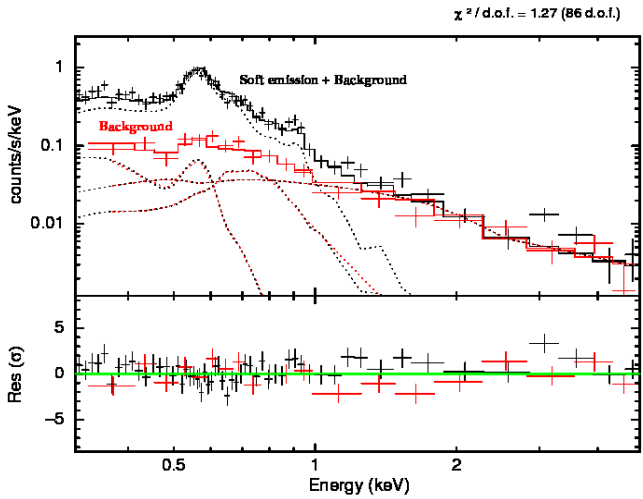}
\caption{XMM-Newton EPIC-pn spectrum and model of the soft emitting region with a VNEI model fit  (black points and solid line). The estimation of the local background spectrum and model (red points and solid line) is described in Section~\ref{subsection:SpectrumBackground}. The background spectrum was not subtracted from the source spectrum but was modelled simultaneously and is included in the spectral model of the source spectrum. The different spectral components are shown separately (red and black dotted lines) for the background and source spectra. The bottom panel shows the residuals from the best-fit models.}
\label{fig:figure2}
\end{center}
\end{figure}

The resulting spectrum is presented in Fig.~\ref{fig:figure2}. The corresponding spectral parameters are given in Table~\ref{tab:Spectrum}. 

\begin{figure}[ht!!]
\begin{center}
\includegraphics[width=0.37\textwidth, angle=-90.0]{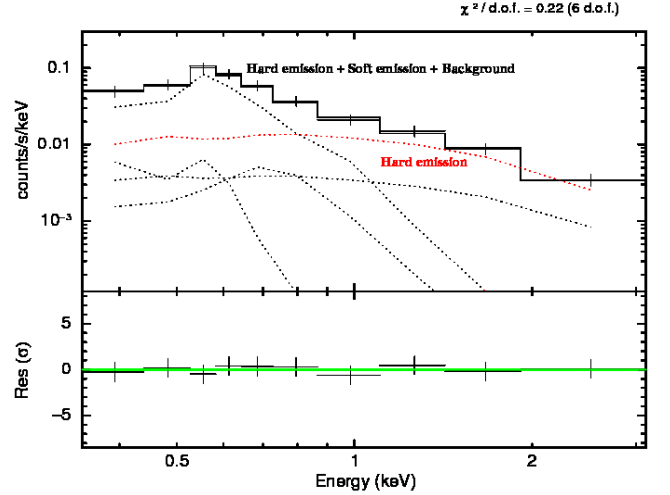}
\caption{XMM-Newton EPIC-pn spectrum of the hard emitting region with a power-law model fit. The solid line is for the hard emission plus background spectra. Here, the background is the sum of the  shell emission and the local emission. The background spectrum and shell spectrum were evaluated as described in Section~\ref{subsection:SpectrumPWN}. The background spectrum was not subtracted from the source spectrum but was modelled simultaneously and is included in the spectral model of the source spectrum. The contribution from the hard emitting region is represented in red. The bottom panel shows the residuals from the best-fit model.}
\label{fig:spectrum2}
\end{center}
\end{figure} 

\subsubsection{Hard emission region}
 \label{subsection:SpectrumPWN}

The X-ray emission above 1.0~keV is dominated by a narrow structure of length $\sim$~0.9\arcmin\ located $\sim$~1\arcmin\ to the west of the centroid of the shell-like thermal emission, as seen from Figure \ref{fig:figure1bis}. Above 1.0 keV, the hard source has a surface brightness of $\sim$~5.8~$\times$~10$^{-3}$ counts/s/arcmin$^2$, i.e. $\sim$~6 times the background level, which has a surface brightness of $\sim$~9.4~$\times$~10$^{-4}$ counts/s/arcmin$^2$. There is indication for substructure in the X-ray image, which could be caused by multiple sources. The \ROSAT\ source \SourceHP{} coincides with the northern part of the narrow structure.

To analyse the spectrum of this second component again the instrumental background spectrum was subtracted and the X-ray spectrum was modelled simultaneously, this time using the surrounding shell-like emission as the local background. The normalisations of the component corresponding to the X-ray background components were renormalised to the area and were fitted, along with the spectral parameters of the hard emitting region.

The spectrum of the hard emitting region is non-thermal and can be modelled in the 0.3~--~5.0 keV energy range with an absorbed power-law with a spectral index of $\Gamma = 1.8 \pm 0.3$, with an absorption of N$_H$~=~0.6~$\times$~10$^{21}$~cm$^{-2}$ for the Galactic foreground (fixed, with solar abundances) and a value of  N$_H$ for the absorption in the LMC of N$_H$~=~(0.58 $\pm$ 0.53)~$\times$~10$^{21}$~cm$^{-2}$. This yields an absorbed flux of $\sim$~$9.5~\times~10^{-14}$~erg/cm$^{2}$/s (reduced $\chi^2$ = 0.22 for 6 degrees of freedom). The resulting spectrum is shown in Fig.~\ref{fig:spectrum2}. In particular, the contribution from the hard X-ray emitting region is represented in red. 

An alternative technique was used to derive the spectrum of the hard emitting region. Instead of using the FWC data, we extracted a background spectrum from a nearby region located within the shell and subtracted it from the source spectrum. This method yields consistent results for the hard emission.

\begin{figure}[t]
\begin{center}
\includegraphics[width=0.39\textwidth, angle=-90.0]{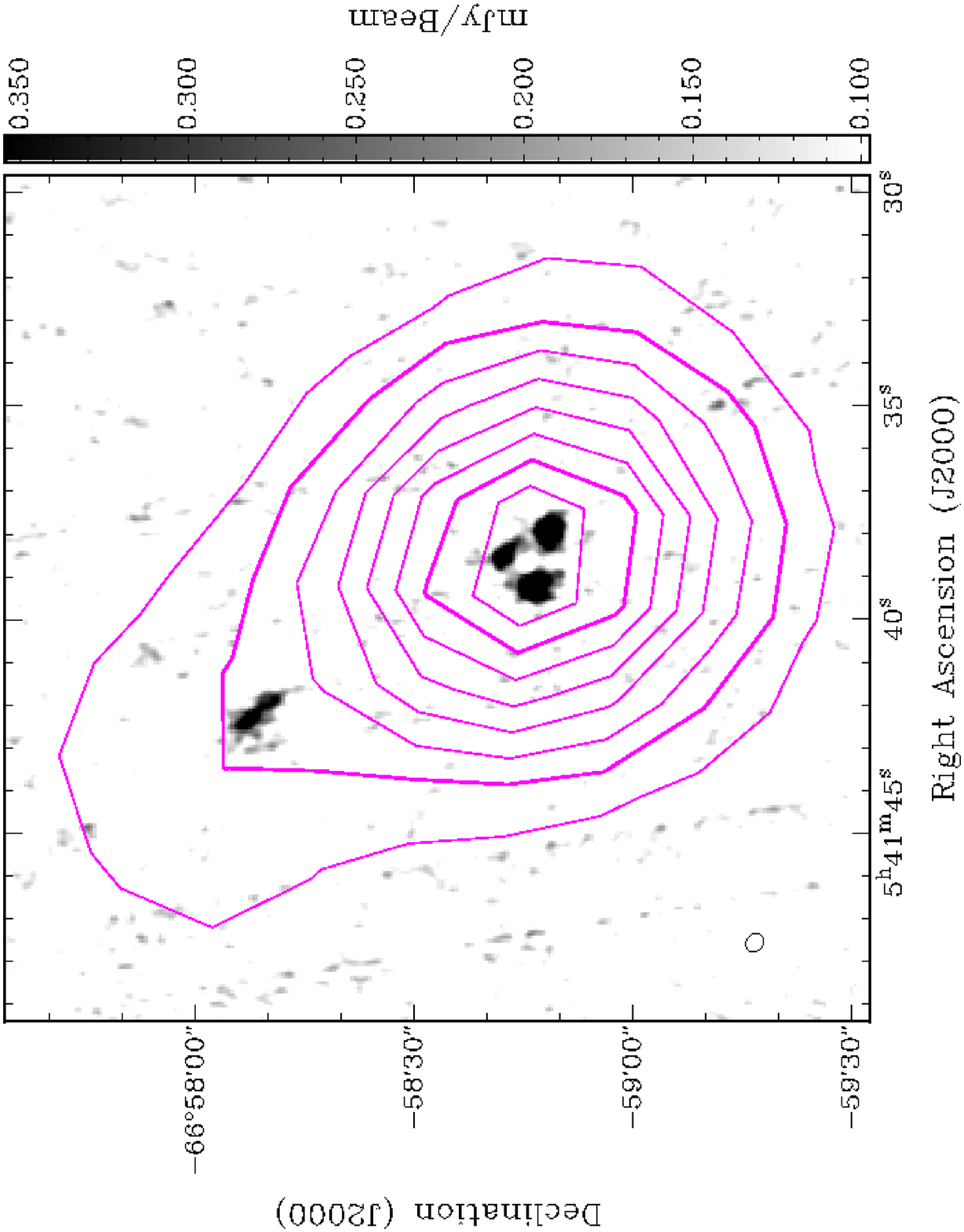}
\caption{High-resolution ATCA radio-continuum image at 6~cm (5.5~GHz) overlaid with 20~cm contours (blue) from the low-resolution mosaic image of the immediate surroundings of \SourceHP{}. Contours are from 3 to 18~mJy/beam in steps of 1~mJy/beam and 6~cm beam size (2.8\arcsec\ $\times$2.2\arcsec) is shown in the bottom left corner.}
\label{fig:figure.radio2}
\end{center}
\end{figure} 

\subsection{Radio}

New high-resolution ($\sim$2\arcsec) radio-continuum observations resolved low-resolution emission into three bright sources within the hard X-ray emitting region (Fig.~\ref{fig:figure.radio2}). For comparison, the size of the radio image shown in Fig.~\ref{fig:figure.radio2} is represented in Fig.~\ref{fig:figure1bis}. In particular, three bright sources can be seen coinciding with the central position of this hard non-thermal X-ray emitting region. The brightest source (the most eastern one) has a spectral index of $\alpha$~=~--0.5~$\pm$~0.1 with integrated flux densities of (0.85$\pm$0.05)~mJy at 5.5~GHz and (0.67$\pm$0.05)~mJy at 9.0~GHz. 

Our estimate of the SNR overall radio spectral index (excluding central point source; using ßux densities from mosaic surveys only and listed in Table~\ref{tab-flux} i.e. observations from 843, 1377, 4800, 8640 MHz) is $\alpha$~=~--1.0~$\pm$~0.2, which is more typical for younger SNRs.

\subsection{Optical}

\begin{figure}[ht!!]
\begin{center}
\includegraphics[width=0.48\textwidth, angle=0.0]{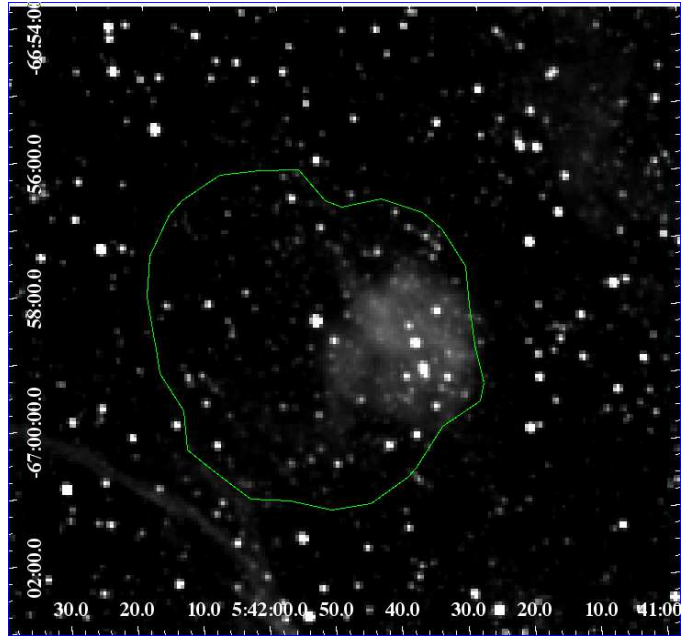}
\caption{\Halpha {} emission (arbitrary units) derived from the MCELS in the surroundings of \SourceHP. A contour of the soft thermal emission of \SourceXMM\ is overlaid in green for comparison.}
\label{fig:figureHa}
\end{center}
\end{figure}

The emission from optical lines (\Halpha\ and \SII) derived from the MCELS were considered. The \Halpha\ emission in the surroundings of \SourceHP{} are presented in Fig.~\ref{fig:figureHa}. This image shows a possible correlation between the western part of the shell-like X-ray thermal emission observed by \XMM, the low-resolution radio-continuum and the \Halpha\ emission line, as seen in Figures \ref{fig:figure.radio2} and~\ref{fig:figureHa}. However, we acknowledge that this correlation could be a chance coincidence because the \Halpha\ ``cloud'' could be an unrelated H{\sc ii} region.

Low-resolution radio-continuum images at 36, 20, 6 and 3~cm show good alignment with the optical (MCELS) feature and also coincide with the maximum emission from the X-rays. 

Supernova remnants can show significant \Halpha, \SII{} and \OIII{} line emission if they are in the radiative phase. In this case, the flux ratio \SII/\Halpha{} is a crucial parameter to distinguish between SNRs and e.g., H{\sc ii} regions \citep{Levenson1995}. However, the \SII/\Halpha\ ratio around \SourceHP{} is $\sim$~0.3, which is not necessarily indicative of SNR emission. 

\Halpha{} emission is spatially coincident with the hard emitting region, but additional studies are required to determine if they are associated or not.

\begin{figure}[h!!]
\begin{center}
\includegraphics[width=0.48\textwidth, angle=0.0]{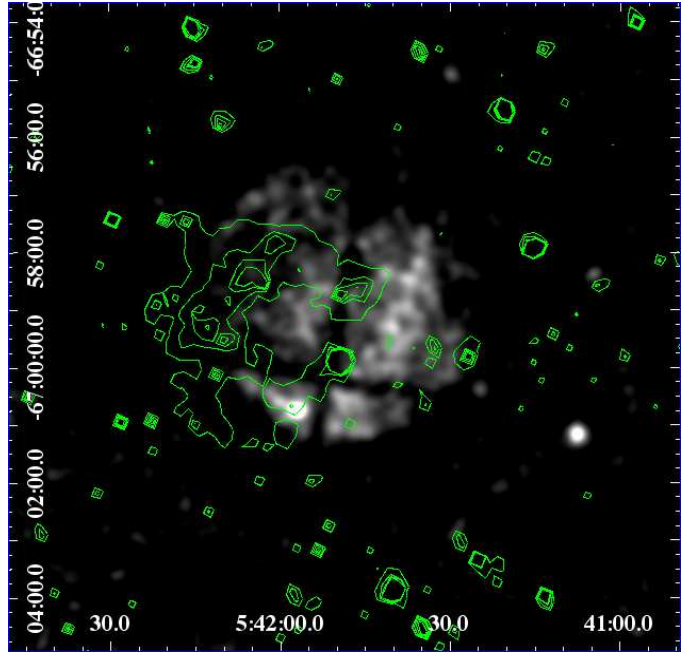}
\caption{Soft X-ray emission of \SourceXMM. The contours of the near-infrared emission ({\it Spitzer}, 5.8 $\mu$m) are overlaid for comparison.}
\label{fig:figureIR}
\end{center}
\end{figure} 

\section{Discussion}
 \label{section:discussion}

Thanks to the high sensitivity of \XMM{}, a recent observation of the field of \SourceHP{} has revealed the existence of two emitting components. The softer component presents a circular morphology and is firmly identified as a new SNR, designated \SourceXMM. The nature of the emitting region with harder spectrum is still unclear, but we examined the possible identification as a PWN or a background AGN.

\subsection{Properties of the newly identified SNR \SourceXMM}
 \label{section:discussionSNR}

X-ray data reveal a large structure of soft emission close to the \ROSAT\ PSPC source \SourceHP{} in the LMC. Analyses performed in different energy ranges indicate that this source dominates the X-ray emission below 1.0~keV.

The spectrum of the SNR \SourceXMM\ can be modelled with an absorbed component representing a plasma in non-equilibrium ionisation characterised by a temperature of $\sim$~0.49 keV, which is the average of the typical temperature of previously detected SNRs in the LMC. We note that the spectral parameters quoted in Table~\ref{tab:Spectrum} are consistent with the parameters of the unambiguously identified SNRs in the LMC \citep{Klimek2010, Levenson1995, Williams1999, Williams2004}. It further supports the identification of this thermal emission with a SNR.

The soft emitting region is firmly identified as a new SNR, in view of its morphology and X-ray spectrum. It has a maximum diameter of $\sim$~5.0\arcmin, which corresponds to a maximum extent of $\sim$~73 pc at a distance of $\sim$~50 kpc. Therefore, \SourceXMM{} is one of the largest SNR observed in the LMC \citep{Williams1999, Klimek2010}. 

However, the large extent of the source does not necessarily imply an old age for the SNR. Indeed, assuming the SNR to be in the Sedov phase, the dynamic age of the source can be estimated using the shock temperature as follows :
\begin{equation}
k_B T_s = 1.8 \times 10^5 \left( \frac{R_s}{t} \right)^2\, \rm{keV} \rm,
\end{equation}
where $k_B$ is the Boltzmann constant, $T_s$ is the temperature of the shock, which is comparable here to the plasma temperature estimated in Section~\ref{subsection:SpectrumShell}, $R_s$ is the radius of the shock (in pc) and $t$ is the age of the SNR (in yr). 

From the spectral fit yielding a plasma temperature of $k_B T_s \approx 0.49$~keV, we derived an age of $\sim$~23~kyr. This value is well below the age of previous largely-extended SNRs detected in the LMC \citep[e.g.][]{Klimek2010, Williams2004}.

This result means that the large extent of the source is not due to its high age, but rather to an expansion in a probably rarefied ambient medium. This is not surprising knowing the position of \SourceXMM{} at the north-east part of the LMC, where the gas density is quite low.

As a legacy project of the {\it Spitzer} Space Telescope, a survey of the Magellanic Clouds has been carried out called ``Surveying the Agents of Galaxy Evolution'' (SAGE)\footnote{{\it Spitzer}-SAGE: http://sage.stsci.edu/index.php} \citep{Meixner2006} to obtain images and spectra of the dust emission. A uniform survey of the LMC was performed in a 7$^{\circ}$~$\times$~7$^{\circ}$ by the Infrared Array Camera (IRAC, dedicated to near infrared observations at wavelength 3.6, 4.5, 5.8, and 8 $\mu$m) and Multiband Imaging Photometer for Spitzer (MIPS, mid- and far-infrared observations at wavelengths 24, 70, and 160 $\mu$m). The near-infrared emission (at 5.8 $\mu$m) in the region surrounding the source \SourceHP{} is presented in Fig.~\ref{fig:figureIR}. {\it Spitzer}-SAGE observations \citep{Meixner2006} reveal dust emission spatially coincident with the eastern half of the shell-like soft emitting region. However, there is no clear indication that this dust emission is associated to the source \SourceXMM. 

\subsection{Hard emission region: a PWN or background AGN?}
 \label{section:discussionPWN}

Energy-dependent morphological studies of the X-ray emission in the surroundings of \SourceHP{} have revealed an extended elongated structure located within the SNR, which dominates the X-ray emission above 1.0~keV. Its centre is located $\sim$~1\arcmin\ from the centre of the SNR. This hard emission region has an apparent extent of $\sim$~0.9\arcmin, corresponding to $\sim$~13~pc at a distance of $\sim$~50 kpc and a non-thermal spectrum. However, the angular resolution of \XMM\ does not allow us to clarify if this hard source is indeed extended (with substructure) or if it is a conglomerate of point sources. A careful look at the \ROSAT\ images reveals that the position of \SourceHP{} is consistent with the hard source detected with \XMM\ and in particular with the more northern knot in the hard emission (see Fig.~\ref{fig:figure1}). Hence, a possible association between the soft and the hard components and the exact nature of the hard X-ray source remains unclear. The radio observations reveal several point sources at the position of the northern X-ray knot. This may point towards an identification as a pulsar with PWN. However, equivalently, one may argue for an existence of several background sources such as AGN. We investigate the different scenarios below.

Firstly, we examine the possibility that the hard X-ray emitting region is associated to background AGN. With a flux of $5.1 \times 10^{-14}$~erg~cm$^{-2}$~s$^{-1}$ (0.5~--~2~keV), we expect to see $\sim$~12 AGN per square degree in the sky \citep{Rosati2002}. We have an 8~\% probability for an AGN inside a 5.0'x4.5' SNR shell. This low probability and  the non-detection of any similar morphological (jet-like) structure either in radio, IR or in optical renders an association of the hard X-ray component with an AGN quite unlikely.

It is worth noticing that the different values of absorption obtained during the spectral fit of the hard and soft emitting regions may indicate that they are located at different distances from us. While the absorption of the shell is consistent with the SNR being located in the LMC, the slightly higher absorption of the hard emission suggests that the source might be located behind the SNR. 

Secondly, we consider the possible association of the hard X-ray component with a PWN. Three radio sources are embedded in the hard and narrow X-ray component. The brightest radio source presents a spectrum that can be modelled with a power-law of index $\alpha \approx -1.0$ which is consistent with spectra of pulsars. If this radio source is indeed a pulsar, then it might power a PWN that would be seen in the X-ray domain. Timing observations are required to confirm or invalidate this identification. Furthermore, it is worth noticing that with a spectral index of $\Gamma = 1.8 \pm 0.3$, the spectrum of the hard X-ray emitting source is similar to spectra of PWNe in the Milky Way \cite[spectral indices ranging from --2.5 to --1.2; ][]{Kargaltsev2008, Kargaltsev2010} but also typical for AGN.

The multi-frequency observations along with the spectral and morphological results of our X-ray and radio data analyses do not uniquely identify the nature of the hard X-ray emission embedded in the SNR \SourceXMM. Deeper X-ray observations with higher angular resolution are required to differentiate between the two scenarios.

\section{Conclusions}
 \label{section:conclusions}

A recent X-ray observation with \XMM{} of \SourceHP{} previously discovered by \ROSAT\ led to the identification of a new SNR (\SourceXMM) in the LMC, which presents a shell-like morphology and a soft thermal spectrum. No clear correlation can be found between the X-ray and optical emission. The physical properties (temperature, size, etc.) of this new SNR are consistent with SNRs previously identified in the LMC by X-ray observations \citep{Levenson1995, Klimek2010, Williams1999, Williams2004}. 

Additional analyses of the \XMM{} data have revealed a harder and narrower emitting region within the shell, which may be a PWN or background AGN. This emission is likely the counterpart of the hard source discovered by \ROSAT. Follow-up observations with ATCA revealed several radio point sources coincident with the hard emitting region, one of which could be a pulsar. Deeper observations of the hard X-ray emitting region with the {\it Chandra} X-ray Observatory will help to distinguish between the different scenarios and to unveil its nature. \\

\noindent{\it Acknowledgements}

The XMM project is supported by the Bundesministerium f$\rm\ddot{u}$r Wirtschaft und Technologie/Deutsches Zentrum f$\rm\ddot{u}$r Luft- und Raumfahrt (BMWI/DLR, FKZ 50 OX 0001) and the Max-Planck Society. 

The Australia Telescope Compact Array is part of the Australia Telescope, which is funded by the Commonwealth of Australia for operation as a National Facility managed by the CSIRO. 

This work is based in part on observations made with the Spitzer Space Telescope, obtained from the NASA/IPAC Infrared Science Archive, both of which are operated by the Jet Propulsion Laboratory, California Institute of Technology under a contract with the National Aeronautics and Space Administration.

The Magellanic Clouds Emission Line Survey (MCELS) data are provided by R.~C. Smith, P.~F. Winkler, and S.~D. Points. The MCELS project has been supported in part by NSF grants AST-9540747 and AST-0307613, and through the generous support of the Dean B. McLaughlin Fund at the University of Michigan, a bequest from the family of Dr. Dean B. McLaughlin in memory of his lasting impact on Astronomy. The National Optical Astronomy Observatory is operated by the Association of Universities for Research in Astronomy Inc. (AURA), under a cooperative agreement with the National Science Foundation.

M.-H. Grondin is supported by the BMBF/DLR grant \#50-0R-1009.

\end{document}